\documentstyle[aps,twocolumn]{revtex}

\input{epsf}

\title{Measurability of Wilson loop operators\thanks{CALT-68-2308}}
\author{David Beckman,$^{(1)}$\thanks{\tt
beckman@caltech.edu} Daniel Gottesman,$^{(2)}$\thanks{\tt gottesma@eecs.berkeley.edu} Alexei Kitaev,$^{(1)}$\thanks{\tt kitaev@iqi.caltech.edu} and John Preskill$^{(1)}$\thanks{\tt
preskill@theory.caltech.edu}}

\address{$^{(1)}$Institute for Quantum Information, California Institute of Technology, 
Pasadena, CA 91125, USA\\
$^{(2)}$Computer Science Division, EECS, University of California, Berkeley, CA 94720, USA\\ }

\begin{document}

\maketitle

\begin{abstract}
We show that the nondemolition measurement of a spacelike Wilson loop operator $W(C)$ is impossible in a relativistic non-Abelian gauge theory. In particular, if two spacelike-separated magnetic flux tubes both link with the loop $C$, then a nondemolition measurement of $W(C)$ would cause electric charge to be transferred from one flux tube to the other, a violation of relativistic causality. A destructive measurement of $W(C)$ is possible in a non-Abelian gauge theory with suitable matter content. In an Abelian gauge theory, many cooperating parties distributed along the loop $C$ can perform a nondemolition measurement of the  Wilson loop operator if they are equipped with a shared entangled ancilla that has been prepared in advance. We also note that Abelian electric charge (but not non-Abelian charge) can be transported superluminally, without any accompanying transmission of information.
\end{abstract}


\section{Introduction and summary}
\label{sec:introduction}

What measurements are possible in a gauge field theory? Since the interactions of the elementary constituents of matter are described by gauge theory, hardly any question could be more fundamental. Yet definitive answers are elusive.

The Wilson loop operators associated with closed spacelike paths provide a complete characterization of a gauge-field configuration in terms of gauge-invariant quantities \cite{giles,gambini}.  Therefore, in formulations of gauge theories, Wilson loops are often taken to be the basic observables. But we will show that nondemolition measurements of spacelike Wilson loops are impossible in a non-Abelian gauge theory that respects relativistic causality. 
We reach this conclusion by arguing that any procedure for nondemolition measurement of a spacelike non-Abelian Wilson loop would allow information to be transmitted outside the forward light cone.

Causality places no such restriction on the measurability of an Abelian Wilson loop (one evaluated in a one-dimensional irreducible representation of the gauge group), and indeed we find that nondemolition measurement of an Abelian Wilson loop is possible. 
We also find that, in gauge theories with suitable matter content, {\em destructive} measurements of non-Abelian Wilson loops are possible. By destructive measurements we mean ones that, in contrast to nondemolition measurements, inflict damage on Wilson loop eigenstates. 

In a quantum field theory in flat spacetime, described in the Schr\"odinger picture, what do we mean by a nondemolition ``measurement'' of an observable defined on a time slice? Typically, such a measurement requires the cooperation of many parties who are distributed over the slice, and is a three-step process. In the first step (which might not be necessary), a suitable entangled quantum state (the ``ancilla'') is prepared and distributed to the parties. Second, each party performs a local operation on her local field variables and her part of the entangled ancilla. Third, classical or quantum information extracted by the parties in the second step is shipped to a central location where the readout of the result is completed.

Although the outcome of the measurement is not known until the third step is completed, the coherence of a superposition of eigenstates of the observable with distinct eigenvalues is already destroyed in the second step, which is carried out on the time slice where the operator is defined. At that time, the density operator $\rho$ encoding the quantum state of the field theory is transformed according to
\begin{equation}
\label{meas_so}
\rho\to {\cal E}(\rho)\equiv\sum_a E_a\rho E_a~,
\end{equation}
where $\{E_a\}$ is the set of orthogonal projectors onto the eigenspaces of the observable. The term ``nondemolition'' means that if the state prior to the measurement is an eigenstate of the observable, then the state will be unaffected by the measurement.

\begin{figure}[t]
\begin{center}
\leavevmode
\epsfxsize=3in
\epsfbox{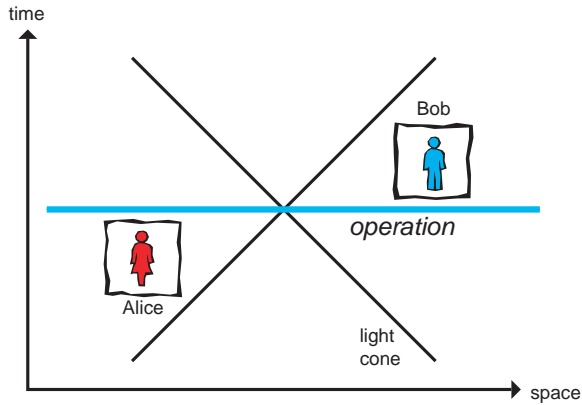}
\end{center}
\caption{Causality criterion for quantum operations. First Alice applies a local operator in her vicinity, then the quantum operation is executed, and finally Bob makes a local measurement that is spacelike separated from Alice's action. If Bob's measurement result allows him to acquire information about what local operator Alice applied, then the quantum operation is not {\em causal} and hence not physically implementable.}
\label{fig:causal}
\end{figure}

Any permissible way in which a quantum state can change is described by a {\em quantum operation}, a completely positive trace-nonincreasing linear map of density operators to density operators \cite{chuang,preskill229}. The orthogonal measurement ${\cal E}$ in eq.~(\ref{meas_so}), summed over its possible outcomes, is a special type of quantum operation.  It is natural to ask, what are the quantum operations that can really be executed on a time slice in a relativistic quantum theory? The general answer is not known, but it {\em is} known that many operations are unphysical because they run afoul of relativistic causality \cite{ahar80,ahar81,ahar86,sorkin,popescu,beckman,nielsen}.  Consider, as in Fig.~\ref{fig:causal}, two parties Alice and Bob who perform spacelike-separated actions. Just prior to the implementation of ${\cal E}$, Alice performs a local operation on the fields in her vicinity, and just after the implementation of ${\cal E}$, Bob performs a local measurement of the fields in his vicinity. If Bob is able to acquire any information about what local operation Alice chose to apply, then Alice has successfully sent a superluminal signal to Bob. If an operation allows such superluminal signaling, we say that the operation is {\em acausal}; otherwise, it is causal. Physically realizable operations must be causal. We will apply this causality criterion to non-Abelian gauge theories, and will argue that nondemolition measurement of a Wilson loop operator is an acausal operation. 

In discussing the locality properties of a field theory, it is convenient to use the concept of a ``reduced'' density operator that encodes the observations that are accessible to an agent acting in a bounded spatial region. This density operator is obtained from a density operator for the full system by ``tracing out'' the degrees of freedom in the unobserved region. In a gauge theory, performing a partial trace involves potential subtleties arising from the Gauss law constraint satisfied by physical states. For conceptual clarity, we will sidestep these difficulties by founding our discussion on the concept of charge super-selection sectors  \cite{haag}. Strictly speaking, our analysis applies to the ``free-charge'' phase of a weakly-coupled gauge theory with non-Abelian gauge group $G$; the local symmetry is unbroken and $G$ charges are unconfined. The same argument, though, shows that Wilson loop measurement would allow superluminal signaling in a confining gauge theory, where the separation between the communicating parties is small compared to the confinement distance scale. 

\begin{figure}[t]
\begin{center}
\leavevmode
\epsfxsize=3in
\epsfbox{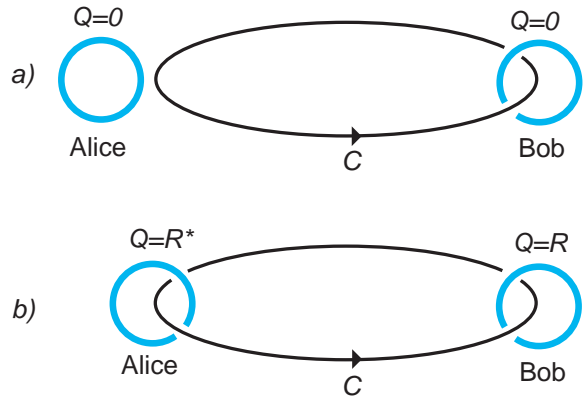}
\end{center}
\caption{Nondemolition measurement of the Wilson loop operator $W(C)$ (more precisely decoherence in the basis of eigenstates of $W(C)$) allows Alice to signal Bob. Alice and Bob, with gauge-invariant probes, can prepare magnetic flux tubes that carry trivial electric charge. In $(a)$, Bob's tube links with the loop $C$ but Alice's does not; when $W(C)$ is measured, neither tube is affected. In $(b)$, Alice moves her tube into position so that it too links with $C$; when $W(C)$ is measured, then (with nonvanishing probability), Bob's tube and Alice's acquire nontrivial and opposite electric charges. By measuring the charge of his tube, Bob can tell how Alice positioned her loop and so receive a message from Alice.}
\label{fig:wilson}
\end{figure}

\begin{figure}
\begin{center}
\leavevmode
\epsfysize=1in
\epsfbox{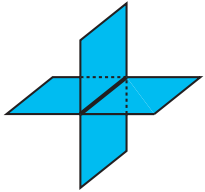}
\end{center}
\caption{A small ``magnetic flux tube'' in lattice gauge theory. By manipulating her link, Alice excites the plaquettes that contain the link, creating a magnetic flux tube. The links dual to these plaquettes form a closed loop on the dual lattice.}
\label{fig:link}
\end{figure}

The protocol by which Alice can exploit measurement of the spacelike Wilson loop operator $W(C)$ to send a signal to Bob is illustrated in Fig.~\ref{fig:wilson}. First Alice and Bob, acting on the weakly-coupled ground state with gauge-invariant local probes, both create {\em magnetic flux tubes}. Bob's flux tube links with the loop $C$; Alice encodes one bit of classical information by placing her tube in one of two possible positions, either linking with $C$ or not.  In the framework of lattice gauge theory, we may imagine that Bob has control of a single lattice link $\ell_B$ contained in the loop $C$, and he creates his ``flux tube'' by manipulating his link --- exciting the lattice plaquettes that contain $\ell_B$ to a particular nontrivial conjugacy class of $G$, as illustrated in Fig.~\ref{fig:link}.  Similarly, Alice controls a single link $\ell_A$ and she encodes a bit by either exciting her link or not. Of course, since Alice and Bob act locally and the theory respects a charge superselection rule, the flux tubes created by Alice and Bob have trivial electric charge.

If either Bob's tube or Alice's tube links with the loop $C$, but not both, then the configuration is an eigenstate of $W(C)$ (or close to an eigenstate in the weakly-coupled case) and will be unaffected (or little affected) by the measurement of $W(C)$. But if both tubes link with $C$, the configuration is not an eigenstate, and will be altered by the measurement. We will see that, with nonvanishing probability, the measurement will generate equal and opposite nonzero electric charges on Alice's tube and Bob's. Then, by measuring the charge on his tube, Bob can infer (with a success probability better than a random guess) whether Alice's tube linked with $C$ or not, and so receive a superluminal signal. 

In a non-Abelian gauge theory, a magnetic flux tube can carry a peculiar kind of electric charge that has no localized source, which has been called Cheshire charge \cite{preskill_krauss,coleman}. (The property that the {\em charge} of an excitation in a non-Abelian gauge theory need not be the integral of a local density is analogous to the property that the {\em energy} of an object in general relativity need not be the integral of a local density.) Our protocol for superluminal signaling is based on the observation that Wilson loop measurement causes Cheshire charge to be transferred from Alice's flux tube to Bob's. Cheshire charge, while conceptually elusive, is physically genuine and readily detected in principle.

Our conclusion that Wilson loop measurement is an acausal operation does not hold in the case of an Abelian gauge theory. Indeed, we will show that a nondemolition measurement of $W(C)$ is possible in an Abelian gauge theory that includes charged matter. In our analysis of this case, we adopt the convenient idealization that the parties who perform the measurement are equipped with gauge-invariant ancilla variables that are not themselves described by the gauge theory. We will also see that a {\em destructive} measurement of a non-Abelian Wilson loop (a measurement that determines the value of the Wilson loop but in doing so damages Wilson loop eigenstates) is possible in a gauge theory that includes suitable charged matter. In particular, if the matter transforms faithfully under the gauge group, then the Wilson loop can be measured destructively in any representation of the gauge group.

The conclusion that nondemolition measurement of spacelike Wilson loops is impossible in a non-Abelian gauge theory seems surprising and somewhat troubling, as it leaves us without a fully satisfactory way to characterize the configurations of a quantized relativistic gauge theory in terms of measurable quantities. Related difficulties arise in quantum theories of gravity. That nondemolition measurement of a Wilson loop operator would allow superluminal signaling has been anticipated by Sorkin \cite{sorkin}.

We formulate the properties of magnetic flux tubes in Sec.~\ref{sec:wilson}, and analyze a protocol for superluminal signaling enabled by Wilson loop measurement in Sec.~\ref{sec:nondemolition}.  In Sec.~\ref{sec:flux}, we defend the legitimacy of the magnetic flux tubes that are used in our signaling protocol, through explicit constructions within the formalism of lattice gauge theory. In Sec.~\ref{sec:matter}, we explore the consequences of including charged matter fields in the gauge theory, and show that destructive measurement of a spacelike Wilson loop is possible. The Abelian case is discussed in Sec.~\ref{sec:abelian}, and we show that in an Abelian gauge theory with charged matter, nondemolition measurement of a Wilson loop is possible. The pure Abelian gauge theory (without matter) is considered in Sec.~\ref{sec:pure}; in that case nondemolition measurement of homologically trivial Wilson loops is possible, but homologically nontrivial Wilson loops are unmeasurable and there is an associated superselection rule. We take up the related question of whether electric charges can travel faster than light in Sec.~\ref{sec:sctl}, concluding that superluminal transport of Abelian charge, but not non-Abelian charge, is possible. Sec.~\ref{sec:conclusions} contains some concluding comments. 

\section{Wilson loops, magnetic flux, and electric charge}
\label{sec:wilson}
In a theory with gauge group $G$, the effect of parallel transport of a charged object around a closed path $C$ that begins and ends at the point $x_0$ can be encoded in a group element $a(C,x_0)\in G$ given by
\begin{equation}
a(C,x_0)= P\exp\left(i\int_{C,x_0} A\right)~.
\end{equation}
Here $A$ is the gauge potential and $P$ denotes path ordering;  the state $|q\rangle$ of a charged object carried along $C$ is modified according to 
\begin{equation}
|q\rangle\to D^{(R)}(a(C,x_0))~|q\rangle~,
\end{equation}
if $|q\rangle$ transforms as the unitary irreducible representation $D^{(R)}$ of $G$. The element $a(C,x_0)\in G$ depends on a ``gauge choice'' at the point $x_0$; that is, on how a basis is chosen in the representation $D^{(R)}$. A basis-independent characterization of the gauge holonomy is obtained if we evaluate the trace in the representation $D^{(R)}$, obtaining the Wilson loop operator associated with $C$ given by
\begin{equation}
W^{(R)}(C)= \chi^{(R)} (a(C,x_0))~,
\end{equation}
where $\chi^{(R)}$ denotes the character of the representation $D^{(R)}$. The Wilson loop operator does not depend on how the point $x_0$ on the loop $C$ is chosen. In much of what follows, we will assume for notational simplicity that the unbroken gauge group $G$ is finite; however, our arguments can be easily extended to the case of compact Lie groups.

By acting with a gauge-invariant source on the weakly-coupled ground state of the gauge theory , Alice (or Bob) can create a ``color magnetic flux tube'' or ``cosmic string'' that carries trivial ``color electric'' charge. This tube is an eigenstate of the Wilson loop operator $W^{(R')}(C_A)$, where $C_A$ is a loop that links once with the tube, for any irreducible representation $(R')$ of the gauge group G; hence the tube can be labeled by a conjugacy class $\alpha$ of $G$. 

When we say that the tube has trivial gauge charge, we mean that it transforms as the trivial representation of $G$ under global gauge transformations. To understand this property it is helpful to specify a basepoint $x_{0,A}$ on the loop $C_A$ and to fix the gauge at this point. Then the effect of parallel transport around the loop $C_A$, beginning and ending at $x_{0,A}$, can be encoded (in this particular gauge) in a group element $a(C_A,x_{0,A})\equiv a$. If the tube is associated with a particular group element $a$, we call its quantum state a ``flux eigenstate,'' denoted $|a\rangle$. But under a gauge transformation $g\in G$ at $x_{0,A}$, this flux eigenstate is transformed as
\begin{equation}
a\to gag^{-1}~.
\end{equation}
Thus a flux eigenstate is not a gauge singlet in general, if $G$ is non-Abelian. A gauge-singlet quantum state of the flux tube is a coherent superposition of the flux eigenstates belonging to conjugacy class $\alpha$,
\begin{equation}
\label{chargezero_alpha}
|\alpha,0\rangle = {1\over \sqrt{|\alpha|}}\left(\sum_{a\in \alpha} |a\rangle\right)~,
\end{equation}
where $|\alpha|$ denotes the number of members of the class. 

Other possible states of the flux tube can carry nontrivial electric charge.  For example, the state
\begin{equation}
|\alpha,R\rangle = N_{\alpha,R}\left(\sum_{g\in G}\chi^{(R)}(g)^*~ |gag^{-1}\rangle\right)~
\end{equation}
(where $N_{\alpha,R}$ is a normalization factor and $a\in \alpha$) transforms as the nontrivial irreducible representation $(R)$ under global gauge transformations. To verify this, first construct the operator
\begin{equation}
E^{(R)}= {n_R\over |G|}\sum_{g\in G} \chi^{(R)}(g) U(g)~,
\end{equation}
where $|G|$ denotes the order of the group, $n_R$ is the dimension of the irreducible representation $(R)$, and $U(g)$ is the global gauge transformation that conjugates the flux by $g$,
\begin{equation}
U(g): |a\rangle \to |gag^{-1}\rangle~.
\end{equation}
Using the group orthogonality relations
\begin{equation}
{n_R\over |G|}\sum_{g\in G} \chi^{(R)}(g)^*\chi^{(R')}(gh)= \delta^{RR'}\chi^{(R)}(h)~,
\end{equation}
we find that $E^{(R)}$ is the orthogonal projection onto the space transforming as $(R)$, which satisfies
\begin{equation}
E^{(R)}E^{(R')}= \delta^{RR'} E^{(R)}~.
\end{equation}
Applying the orthogonality relations once more, we see that
\begin{equation}
E^{(R)}|\alpha,R\rangle= |\alpha,R\rangle~;
\end{equation}
thus $|\alpha,R\rangle$ transforms as $(R)$.
The ``Cheshire charge'' carried by the flux tube in this state can be detected through, for example, the Aharonov-Bohm interactions of the tube with other, distant, flux tubes \cite{preskill_krauss,coleman}.

\begin{figure}
\begin{center}
\leavevmode
\epsfxsize=3in
\epsfbox{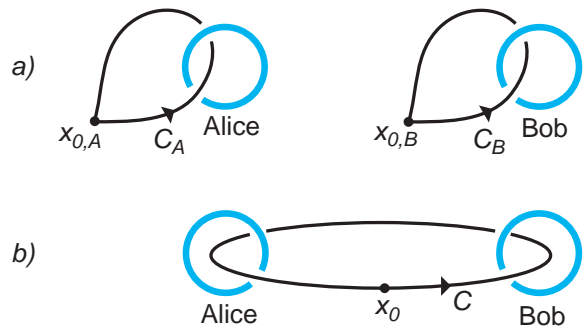}
\end{center}
\caption{Loops linked with flux tubes. The flux tube configuration created locally by Alice and by Bob is an eigenstate of the Wilson loop operators associated with the paths $C_A$ and $C_B$, shown in $(a)$, but not an eigenstate of the Wilson loop operator $C$ that links with both tubes, shown in $(b)$.}
\label{fig:loops}
\end{figure}

Suppose that Alice and Bob, acting locally, each create flux tubes with zero electric charge,  where the flux of Alice's tube belongs to conjugacy class $\alpha$ and the flux of Bob's belongs to conjugacy class $\beta$. (The process of preparing the flux tubes will be discussed in more detail in Sec.~\ref{sec:flux}.) To describe the quantum state of this configuration, we may choose loops $C_A$ and $C_B$ that link with the tubes, and fix the gauge at basepoints $x_{0,A}$ and $x_{0,B}$ as illustrated in Fig.~\ref{fig:loops}a.  Up to a normalization factor, the quantum state of the two tubes can be expressed as
\begin{equation}
\left(\sum_{h\in G} |hah^{-1}\rangle_A\right)\otimes \left(\sum_{g\in G}|gbg^{-1}\rangle_B\right)~,
\end{equation}
where $a\in \alpha$ and $b\in \beta$. This configuration is a direct product of the state of Alice's tube with the state of Bob's tube, a simultaneous eigenstate of the commuting Wilson loop operators $W(C_A)$ and $W(C_B)$. 

But because multiplication of conjugacy classes is ill-defined, this state is not an eigenstate of $W(C)$, where $C$ is the loop shown in Fig.~\ref{fig:loops}b that links with both tubes. Rather, if Alice's tube is an eigenstate of $W(C_A)$ and Bob's tube is an eigenstate of $W(C_B)$, then an eigenstate of $W(C)$ is not a product state but an entangled state of the form (up to normalization)
\begin{equation}
\label{correlated_flux}
\sum_{g\in G} |gag^{-1}\rangle_A\otimes |gbg^{-1}\rangle_B~.
\end{equation}
This state has zero total charge, as it is invariant under a global gauge transformation applied to both tubes.  But it is not invariant under a gauge transformation that acts on just one of the tubes; it can be expanded in a basis in which Alice's tube and Bob's have definite and opposite charges. Using the group orthogonality relations in the form
\begin{equation}
\sum_R {n_R\over |G|} ~\chi^{(R)}(g)=\delta_{g,e}
\end{equation}
(where $e$ is the identity element), we may rewrite the state as a sum over irreducible representations
\begin{equation}
\sum_R{n_R\over |G|}\left(\sum_{h,g\in G}\chi^{(R)}(gh^{-1})^* ~|hah^{-1}\rangle_A\otimes |gbg^{-1}\rangle_B\right)~.
\end{equation}
The expression in parentheses transforms as $(R)$ under gauge transformations acting on Bob's tube and as the conjugate representation $(R^*)$ under gauge transformations acting on Alice's tube, as we can verify by applying $I\otimes E^{(R)}$ and $E^{(R^*)}\otimes I$.

Thus, when two flux tubes are prepared in a quantum state that is an eigenstate of the Wilson loop operator $W(C)$, where $C$ links with both flux tubes, then the flux tubes carry correlated nontrivial electric charges. This property is the basis of our claim that Wilson loop measurement is  acausal, as we elaborate in the next section.

\section{nondemolition measurement of non-Abelian Wilson loops is acausal}
\label{sec:nondemolition}

For a {\em static} gauge field configuration, it is possible in principle to measure $W^{(R)}(C)$ (and hence the conjugacy class of the group element $a(C,x_0)$) by performing interference experiments with projectiles that transform as $D^{(R)}$ \cite{alford}. But what if the loop $C$ lies in a time slice and the gauge field is {\em dynamical}?  In a relativistic field theory, no projectile can follow a spacelike world line, so that a direct measurement of the effect of parallel transport along $C$ is not feasible.

However, it seems conceivable that a less direct measurement strategy might succeed. When we speak of a ``measurement'' of an operator whose support is on a spacelike slice, we need not require that the result be instantaneously known by anyone. We might imagine instead that, in order to measure $W^{(R)}(C)$ at time $t=0$, many parties distributed along the loop $C$ perform local operations at $t=0$. Later, the data collected by the parties can be assembled and processed at a central laboratory, where the outcome of the measurement of $W^{(R)}(C)$ can be determined. In such a protocol, we should allow the parties to share any entangled quantum state that they might have prepared prior to $t=0$, and we should allow them to ship quantum information (rather than just classical data) to the central laboratory after they have performed their local operations. Of course, the quantum or classical variables that are sent to the central laboratory for analysis are not the variables of the underlying field theory; they are ancilla variables that are assumed to be available to assist with the measurement.

Just prior to the measurement at time $t=0$, the quantum state of the gauge theory is $\rho$, a density operator that acts on the physical gauge-invariant subspace. Even though the measurement result may not be known until later, the operations performed at $t=0$ modify the state $\rho$ immediately. If the local operations performed at $t=0$ are to achieve a measurement of $W^{(R)}(C)$, then the coherence of a superposition of eigenstates of $\rho$ with different eigenvalues must be destroyed. At $t=0$, then, the quantum state is modified according to
\begin{equation}
\label{wilson_decohere}
\rho \to {\cal E}_{W(C)}(\rho)\equiv \sum_w E_w \rho E_w~,
\end{equation}
where $E_w$ is the orthogonal projector onto the subspace of states with
\begin{equation}
W^{(R)}(C)= w~.
\end{equation}
This operation describes a projective measurement of $W^{(R)}(C)$ with an unknown outcome.

The operation ${\cal E}_{W(C)}$ is actually weaker than a measurement of $W^{(R)}(C)$; conceivably decoherence in the basis of eigenstates of an observable can be accomplished even if the measurement outcome is {\em never} recorded. But if any record of the value of $W^{(R)}(C)$ is written at $t=0$ (even one that cannot be read until later), then decoherence as described by eq.~(\ref{wilson_decohere}) must occur.

We will show that ${\cal E}_{W(C)}$ can be used to send superluminal signals, and so establish that ${\cal E}_{W(C)}$ cannot be implemented in a gauge theory that respects relativistic causality. 

To devise a superluminal signaling protocol, Alice and Bob use local gauge-invariant probes to prepare uncharged flux tubes belonging to classes $\alpha$ and $\beta$ respectively, as described in Sec.~\ref{sec:wilson}. Bob moves his flux tube, which will receive the message, into position so that it links once with the loop $C$; Alice encodes one bit of information by choosing to place her flux tube in one of two possible positions, either linking with $C$ or not. If Alice chooses to place her tube where it does not link with $C$, then the configuration is an eigenstate of $W^{(R)}(C)$ and will be unaffected when the Wilson loop is measured. But if Alice moves her flux tube into position to link with $C$, then the configuration is no longer an eigenstate of $W^{(R)}(C)$, and it {\em is} affected by the operation ${\cal E}_{W(C)}$. In fact, after the operation, though the total charge of the system remains zero, there is a nonzero probability that Alice's tube and Bob's tube carry equal and opposite nonzero charges. This charge can be detected by Bob. For example, he can determine whether his tube has vacuum quantum numbers by allowing it to shrink and observing whether it will annihilate and disappear --- if the tube is charged, a stable charged particle will be left behind. 

Thus, if there were a way to implement the operation ${\cal E}_{W(C)}$ at $t=0$, then by observing whether his flux tube is charged after $t=0$, Bob would be able to infer (with a probability of success better than a random guess) whether Alice moved her tube into position or not. Therefore Alice can transmit classical information to Bob over a noisy channel with nonzero capacity; she is able to send a superluminal signal to Bob. By the same method, Bob can send a signal to Alice.

To understand this charge transfer process in more detail, let us consider a specific example. Suppose that $G$ is the  {\em quaternionic group} of order eight, whose two-dimensional faithful unitary irreducible representation is 
\begin{equation}
\{\pm I, \pm i\sigma_1,\pm i\sigma_2,\pm i\sigma_3\}~,
\end{equation}
where $\sigma_1,\sigma_2,\sigma_3$ are the Pauli matrices. Suppose that Alice and Bob both have tubes carrying flux in the class $\alpha=\beta=\{\pm i\sigma_1\}$. For tubes in this class, the quantum state with trivial charge is
\begin{equation}
|+\rangle = {1\over \sqrt{2}}\left(|i\sigma_1\rangle + |-i\sigma_1\rangle\right)~,
\end{equation}
and there is also a state of nontrivial charge
\begin{equation}
|-\rangle = {1\over \sqrt{2}}\left(|i\sigma_1\rangle - |-i\sigma_1\rangle\right)~.
\end{equation}
The state $|-\rangle$ transforms as the nontrivial one-dimensional representation of $G$ in which $\pm I$, $\pm i\sigma_1$ are represented by 1 and $\pm i\sigma_2$, $\pm i\sigma_3$ are represented by $-1$.

If Alice and Bob each have a charge-zero flux tube, the quantum state of their two tubes is a product state
\begin{equation}
|\psi\rangle_{\rm init}= |+\rangle_A\otimes |+\rangle_B~.
\end{equation}
But if the loop $C$ links once with each tube, then the value of $W^{(R)}(C)$ in the two-dimensional irreducible representation $(R)$ can be either $2$ or $-2$. If the initial state $|\psi\rangle_{\rm init}$ is projected onto the state with $W^{(R)}(C)=2$, Alice's tube becomes entangled with Bob's; the resulting state is 
\begin{eqnarray}
|\psi\rangle_{{\rm fin}, 2}= {1\over\sqrt{2}}( & &|i\sigma_1\rangle_A\otimes |-i\sigma_1\rangle_B\nonumber\\
& &+ |-i\sigma_1\rangle_A\otimes |i\sigma_1\rangle_B)~,
\end{eqnarray}
Bob's final density operator, obtained by tracing over the state of Alice's tube, is
\begin{equation}
\rho_{B,{\rm fin}, 2}= {1\over 2}\left ( | i\sigma_1\rangle \langle i\sigma_1| + 
| -i\sigma_1\rangle \langle -i\sigma_1|\right)~,
\end{equation} 
an {\em incoherent mixture} of the two flux eigenstates. Similarly, if $|\psi\rangle_{\rm init}$ is projected onto the state with $W^{(R)}(C)=-2$, the resulting state is 
\begin{eqnarray}
|\psi\rangle_{{\rm fin}, -2}= {1\over\sqrt{2}}( & &|i\sigma_1\rangle_A\otimes |i\sigma_1\rangle_B\nonumber\\
& &+ |-i\sigma_1\rangle_A\otimes |-i\sigma_1\rangle_B)~,
\end{eqnarray}
and again Bob's final density operator is
\begin{equation}
\rho_{B,{\rm fin}, -2}=\rho_{B,{\rm fin}, 2}~.
\end{equation} 
Each of the two flux eigenstates is an equally weighted coherent superposition of the charge eigenstates $|+\rangle_B$ and $|-\rangle_B$. Thus if Bob were to measure the charge of his tube after the operation ${\cal E}_{W(C)}$ acts at $t=0$, he would find the charge to be $(-)$ with probability 1/2 if Alice's tube linked with the loop $C$, while he would never find the charge to be $(-)$ if Alice's tube did not link with $C$. Alice has sent a superluminal signal to Bob.

We can easily generalize this construction to an arbitrary finite gauge group $G$. If Alice's tube initially carries flux in the conjugacy class $\alpha$ and has trivial charge,  while Bob's carries flux in the class $\beta$ and has trivial charge, then the initial state of their tubes is a product state
\begin{equation}
|\psi\rangle_{\rm init} = {1\over\sqrt{|\alpha|\cdot|\beta|}}\sum_{a\in\alpha}\sum_{b\in\beta}|a\rangle_A\otimes |b\rangle_B~.
\end{equation}
This state can be expanded in terms of eigenstates of 
\begin{equation}
W^{(R)}(C)= \chi^{(R)}(ab)~.
\end{equation}
Suppose that for fixed $a\in\alpha$ there are $n_i$ distinct elements $b^{(a)}_{i,\mu}\in \beta~,\mu=1,2,\dots, n_i$, such that $\chi^{(R)}(ab)=w_i$. (This number $n_i$ is independent of how the class representative $a$ is chosen.) Then the component of $|\psi\rangle_{\rm init}$ with $W^{(R)}(C)=w_i$ is the (unnormalized) entangled state
\begin{equation}
|\psi\rangle_{w_i}= {1\over\sqrt{|\alpha|\cdot|\beta|}}\sum_{a\in\alpha}|a\rangle_A\otimes \sum_{\mu=1}^{n_i}|b^{(a)}_{i,\mu}\rangle_B~.
\end{equation}
This state is invariant under a global gauge transformation acting as
\begin{equation}
a\to gag^{-1}~, \quad b\to gbg^{-1}~,
\end{equation}
so that its total charge is trivial.
We see that if $W^{(R)}(C)$ is measured, the outcome $w_i$ occurs with probability
\begin{equation}
{\rm Prob}(w_i)= {}_{w_i}\langle\psi|\psi\rangle_{w_i}= {n_i/|\beta|}~;
\end{equation}
it is obvious from the definition of $n_i$ that these probabilities sum to unity.
Furthermore, if the state $|\psi\rangle_{w_i}$ is prepared by the measurement of the Wilson loop, and Bob subsequently measures the charge of his tube, he will find the charge to be trivial with probability
\begin{eqnarray}
{\rm Prob}(0|w_i)&=& {{}_{w_i}\langle\psi|\left(I_A\otimes |\beta,0\rangle\langle\beta,0|\right)|\psi\rangle_{w_i}\over {}_{w_i}\langle\psi|\psi\rangle_{w_i}}\nonumber\\
&=& n_i/|\beta|= {\rm Prob}(w_i)~. 
\end{eqnarray}
(Here $|\beta,0\rangle$ denotes the charge-0 state of a string whose flux is in conjugacy class $\beta$.) Therefore, if Alice's and Bob's tubes both link once with the loop $C$ when the operation ${\cal E}_{W(C)}$ is applied, then afterwards Bob will find his tube carries trivial charge with probability
\begin{eqnarray}
{\rm Prob}(0)& = &\sum_i {\rm Prob}(0|w_i)\cdot {\rm Prob}(w_i) \nonumber\\
& = &\sum_i ({\rm Prob}(w_i))^2= \sum_i \left(n_i/|\beta|\right)^2~.
\end{eqnarray}
We see that, unless the initial configuration is an eigenstate of $W^{(R)}(C)$, we have ${\rm Prob}(0) <1$. We conclude that Bob's tube is charged with nonzero probability if Alice's tube linked with $C$, and it is guaranteed to be uncharged if Alice's tube did not link with $C$. Alice can send a superluminal signal to Bob. (Of course, since Alice's tube has an electric charge equal and opposite to that of Bob's tube, Bob can also send a superluminal signal to Alice, with the same probability of success.)

The argument also applies to compact Lie groups. For example if the gauge group is
\begin{eqnarray}
G= SU(2)= 
\{g(\hat n ,\theta)=\exp&&\left[-i(\theta/2)\hat n\cdot\vec\sigma\right]~,\nonumber\\
&& \hat n\in S^2~, \quad \theta\in[0,2\pi]\}~,
\end{eqnarray}
then conjugacy classes are labeled by $\theta$. If a flux tube has trivial charge, its quantum state can be expressed as
\begin{equation}
|0,f\rangle = \int d\theta f(\theta)\int d\hat n |g(\hat n,\theta)\rangle~,
\end{equation}
where $f$ is any square integrable class function. If the loop $C$ links with Alice's tube and Bob's, then the product state
\begin{equation}
|\psi\rangle_{\rm init}= |0,f_1\rangle_A\otimes |0,f_2\rangle_B
\end{equation}
is in general not an eigenstate of the operator $W^{(R)}(C)$; hence measurement of $W^{(R)}(C)$ would induce a detectable transfer of charge from Alice's tube to Bob's.
 
The argument also applies in any spatial dimension $d\ge 2$. In $d=2$ dimensions, the flux tubes may be replaced by pairs of pointlike vortices; in $d > 3$ dimensions, the tubes become membranes of codimension 2.

In our discussion, we have ignored the effects of magnetic and electric quantum fluctuations --- in particular we have not considered whether gauge charges might be confined or screened by the Higgs mechanism. We have implicitly assumed that the $G$ gauge symmetry is unbroken, and (if $G$ electric charges are confined) that the separation between Alice and Bob is small compared to the characteristic distance scale of electric confinement. 

We should note that in the case of a continuous gauge group, an ultraviolet regulator is implicitly invoked to define the Wilson loop. The Wilson loop detects the magnetic flux that links with $C$. If we think of $C$ as a wire of infinitesimal thickness, then $W(C)$ will be dominated by very-short-wavelength fluctuations of the gauge field near the wire. To suppress these fluctuations, we allow the wire to have a nonzero thickness $a$, removing the contributions of fluctuations with wavelength below $a$. In $3+1$ spacetime dimensions, the fluctuations near the wire are unimportant provided that 
\begin{equation}
e^2\log(L/a)\ll 1~,
\end{equation}
where $e^2$ is the gauge coupling constant (renormalized at distance scale $a$) and $L$ is the characteristic size of the loop $C$.

\section{Flux tubes on the lattice}
\label{sec:flux}

Our argument that Wilson loop measurement would allow Alice to send a superluminal signal to Bob had two crucial elements: that Alice and Bob are capable of creating uncharged magnetic flux tubes, and that Bob can detect the charge on his tube. Let us examine more deeply whether the preparation of the flux tube is really possible in principle. 

In considering whether flux tubes are legitimate objects, it is helpful to think about a scenario in which an underlying continuous gauge symmetry is spontaneously broken to a finite non-Abelian subgroup. To be specific, a generic vacuum expectation value of a Higgs field in the five-dimensional irreducible representation of $SU(2)$ breaks the gauge symmetry to the quaternionic group considered in Sec.~\ref{sec:nondemolition}. In this Higgs phase, there are locally stable cosmic strings that carry nontrivial magnetic flux; these serve as the flux tubes needed for the signaling protocol.
Alice and Bob both require closed loops of string that have vacuum quantum numbers; in principle, these could be created in, for example, a hard collision between particles.

Wilson loop measurement can change the transformation properties of a string loop under global gauge transformations --- it transfers charge to the loop. This charge, like any charge in a discrete gauge theory, can be detected through the Aharonov-Bohm interactions of the string loop with other string loops \cite{preskill_krauss,coleman}. 

In a confining gauge theory like quantum chromodynamics, a flux tube is not locally stable, but it is still possible to engineer one, at least if it is small compared to the confinement distance scale. To be as concrete as possible, we will describe how a flux tube can be created in a gauge theory defined on a spatial lattice (but with continuous time). In this framework, Bob (or Alice) can prepare a flux tube with zero charge by acting on a single link variable with a gauge-invariant local operator, as indicated in Fig.~\ref{fig:link}. 

In our description of the construction of this operator, we will again find it convenient to suppose that the gauge group $G$ is a finite group of order $|G|$, though there are no serious obstacles to generalizing the discussion to the case of Lie groups. Residing on the lattice links are variables that take values in the $G$ group algebra, a Hilbert space of dimension $|G|$ for which an orthonormal basis can be chosen as $\{|g\rangle, g\in G\}$. A local gauge transformation associates a group element with each lattice site. Each link has an orientation, and if a link connecting sites $x$ and $y$ is oriented so that it points from $y$ to $x$, then gauge transformations $U_x(h)$ and $U_y(k)$ at site $x$ and $y$ act on the link variable according to 
\begin{eqnarray}
&&U_x(h):|g\rangle_{xy}\to |hg\rangle_{xy}~,\nonumber\\
&&U_y(k):|g\rangle_{xy}\to |gk^{-1}\rangle_{xy}~.
\end{eqnarray}
{\em Physical states} are invariant under all local gauge transformations. Physical observables preserve the space of physical states, and hence must commute with the local gauge transformations.

Now consider an operator $H_\ell(a)$ that acts on a particular link $\ell$ as
\begin{equation}
H_\ell(a): |g\rangle_\ell\to |ag\rangle_\ell~.
\end{equation}
Note that $H_\ell(a)$ is not a gauge transformation, since it acts only on a single link, rather than all of the links that meet at a site.
This operator does not commute with local gauge transformations; rather if $\ell$ is oriented so that it points toward the site $x$, we have
\begin{equation}
U_x(h)H_\ell(a)U_x(h)^{-1}:|g\rangle_\ell\to |hah^{-1}\cdot g\rangle_\ell~,
\end{equation}
or 
\begin{equation}
U_x(h)H_\ell(a)U_x(h)^{-1}= H_\ell(hah^{-1})~.
\end{equation}
But if we define an operator by summing $H_\ell(a)$ over a conjugacy class of $G$,
\begin{equation}
H_\ell(\alpha)= {1\over |\alpha|}\left(\sum_{a\in \alpha} H_\ell(a)\right)~,
\end{equation}
then $H_\ell(\alpha)$ {\em does} commute with $U_x(h)$ and is therefore a gauge-invariant operator. This is the operator that Alice applies to her link to create a local flux tube excitation \cite{lee,kitaev}. Of course, by acting on several adjacent links, Alice can create a larger flux tube if she wishes.

If Alice applies this operator to her link and Bob applies it to his link, then the state they prepare (acting on the weak-coupling vacuum) is not an eigenstate of the Wilson loop operator $W(C)$, where $C$ contains both links. There {\em is} an eigenstate of $W(C)$ in which Alice's link is excited to conjugacy class $\alpha$ and Bob's to class $\beta$, and of course this state can be created by a gauge-invariant operator acting on the perturbative vacuum. But the operator cannot be local, since it creates charges on Alice's link and Bob's. It is instructive to construct the nonlocal gauge-invariant operator that creates this state.

\begin{figure}
\begin{center}
\leavevmode
\epsfxsize=3in
\epsfbox{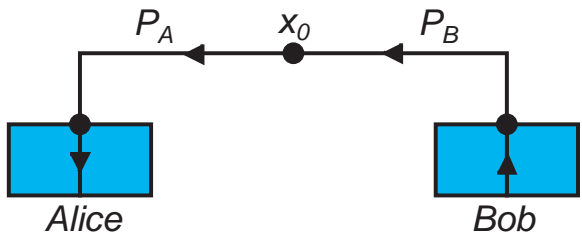}
\end{center}
\caption{A nonlocal operator that creates correlated excitations at distantly separated links of a lattice gauge theory. An arbitrary basepoint $x_0$ is chosen, together with arbitrary paths $P_A$ and $P_B$ that connect Alice's link and Bob's link to the basepoint. By acting on the links, the operator excites the lattice plaquettes (shaded) that contain the links. The nonlocality is necessary because Alice's excitation and Bob's excitation carry nontrivial and opposite electric charges.}
\label{fig:nonlocal_op}
\end{figure}

For this purpose, it is convenient to choose a basepoint lattice site $x_0$, and to choose oriented lattice paths $P_A$ and $P_B$ that connect Alice's link $A$ and Bob's link $B$ to the basepoint, as shown in Fig.~\ref{fig:nonlocal_op}. Let $g_P$ denote the path-ordered product of link variables associated with the path $P$,
\begin{equation}
g_P=\prod_{\ell\in P} g_\ell~,
\end{equation}
with later links along the path appearing further to the left. Then we may define a generalization of the operator $H_\ell$ that depends on the path and the basepoint. Acting on Alice's link $\ell_A$ we have
\begin{equation}
H_{\ell_A}(a,P_A,x_0): |g\rangle_{\ell_A}\to |g \cdot g_{P_A}a g_{P_A}^{-1}\rangle_{\ell_A}~,
\end{equation}
and acting on Bob's link $\ell_B$ we have
\begin{equation}
H_{\ell_B}(b,P_B,x_0): |g\rangle_{\ell_B}\to |g_{P_B}^{-1} b g_{P_B}\cdot g\rangle_{\ell_B}~.
\end{equation}
Hence $H_{\ell_B}(b,P_B,x_0)$, like $H_{\ell_B}(b)$, excites the plaquettes that contain Bob's link. But while $H_{\ell_B}(b)$ left-multiplies the link variable by $b$, $H_{\ell_B}(b,P_B,x_0)$ left-multiplies by the conjugate group element $g_{P_B}^{-1} b g_{P_B}$. In a fixed gauge, the operator $H_{\ell_B}(b,P_B,x_0)$ creates an excitation such that the effect of gauge parallel transport about a closed path that begins and ends at $x_0$ and passes through link ${\ell_B}$ is encoded in the group element $b$.
$H_{\ell_A}(a,P_A,x_0)$ is defined similarly, but acts by right multiplication because of the way we have chosen the orientation of the link ${\ell_A}$.

The operator $H_{\ell_A}(a,P_A,x_0)$ commutes with local gauge transformations acting in the vicinity of Alice's link ${\ell_A}$, and $H_{\ell_B}(b,P_B,x_0)$ commutes with gauge transformations acting in the vicinity of Bob's link ${\ell_B}$. But they do not commute with gauge transformations acting at the basepoint $x_0$; rather we have
\begin{eqnarray}
&&U_{x_0}(g)H_{\ell_A}(a,P_A,x_0)U_{x_0}(g)^{-1}= H_{\ell_A}(gag^{-1},P_A,x_0)~,\nonumber\\
&&U_{x_0}(g)H_{\ell_B}(b,P_B,x_0)U_{x_0}(g)^{-1}= H_{\ell_B}(gbg^{-1},P_B,x_0)~.\nonumber\\
&&
\end{eqnarray}
Again, we can obtain a gauge-invariant operator by summing $a$ or $b$ over a conjugacy class, {\it e.g.},
\begin{equation}
H_{\ell_B}(\beta,P_B,x_0)= {1\over |\beta|}\left(\sum_{b\in \beta} H_{\ell_B}(b,P_B,x_0)\right)~.
\end{equation}
In fact, it is clear from the definitions that $H_{\ell_B}(\beta,P_B,x_0)=H_{\ell_B}(\beta)$; it is really a local operator in disguise.

But we can also construct a gauge-invariant operator that acts simultaneously on Alice's link and Bob's, and that really is nonlocal \cite{lee,kitaev}:
\begin{equation}
\label{correlated_flux_op}
{1\over |G|}\left(\sum_{g\in G} H_{\ell_A}(gag^{-1},P_A,x_0)\cdot H_{\ell_B}(gbg^{-1},P_B,x_0)\right)~.
\end{equation}
This operator, acting on the weak-coupling vacuum, creates a state in which Alice's link and Bob's are correlated, as in eq.~(\ref{correlated_flux}). This gauge-invariant operator does not depend on how the basepoint $x_0$ is chosen; we are free to slide the basepoint along the path connecting Alice's link and Bob's however we please.

The communication protocol explained in Sec.~\ref{sec:nondemolition} can be described this way: Alice and Bob apply the local operators $H_{\ell_A}(\alpha)$ and $H_{\ell_B}(\beta)$ to create link excitations that are uncorrelated with one another. Then the Wilson loop measurement operation ${\cal E}_{W(C)}$ is applied, where the loop $C$ contains the links $\ell_A$ and $\ell_B$. This operation establishes a correlation between the links. It transforms a state that can be created by local operators to a state that can be created only by a nonlocal operator like that in eq.~(\ref{correlated_flux_op}). Such a transformation cannot occur on a time slice in a theory that respects relativistic causality. We conclude that the nondemolition measurement of the non-Abelian Wilson loop operator is not physically realizable.

Now, the operator $H_{\ell}(\alpha)$ is a gauge-invariant local operator, but it is not unitary, so we should clarify what it means to say that Alice or Bob applies this operator to a state.
In fact, if $A$ is any bounded operator that does not annihilate the state $|\psi\rangle$, we can apply the operation
\begin{equation}
\label{A_op}
|\psi\rangle \to {A|\psi\rangle\over \sqrt{\langle\psi|A^\dagger A|\psi\rangle}}
\end {equation}
with a nonzero probability of success by making a suitable measurement. First note that we may assume without loss of generality that the eigenvalues of $A^\dagger A$ are no larger than one --- if not, we merely rescale $A$ without modifying the operation eq.~(\ref{A_op}). Then let $\{|0\rangle,|1\rangle\}$ be an orthonormal basis for a two-dimensional ``ancilla'' space, and consider the transformation
\begin{equation}
U:|0\rangle\otimes |\psi\rangle\to |0\rangle\otimes A|\psi\rangle + |1\rangle\otimes B|\psi\rangle~,
\end{equation}
where 
\begin{equation}
A^\dagger A + B^\dagger B=I ~.
\end{equation}
This transformation is norm-preserving and so has a unitary extension. Hence we apply the unitary $U$ to $|0\rangle\otimes |\psi\rangle$ and then measure the ancilla by projecting onto the basis $\{|0\rangle,|1\rangle\}$. The outcome $|0\rangle$ is  obtained with probability $\langle \psi|A^\dagger A|\psi\rangle$, in which case eq.~(\ref{A_op}) is applied. If the outcome $|1\rangle$ is found, then Alice may discard the state and make another attempt. As long as $A|\psi\rangle\ne 0$, Alice can repeat the procedure until she gets the desired outcome.

Clearly, the gauge-invariant bounded operator analogous to $H_\ell(\alpha)$ can also be constructed in the case where $G$ is a Lie group. For example if the gauge group is $SU(2)$, then associated with the group element $g(\hat n,\theta)=\exp\left(-i(\theta/2)\hat n\cdot \vec\sigma\right)$ is the transformation
\begin{equation}
H_\ell\left(g(\hat n,\theta)\right):|h\rangle_\ell\to |g(\hat n,\theta)h\rangle_\ell~,
\end{equation}
where $h\in SU(2)$. This transformation can be expressed as
\begin{equation}
H_\ell\left(g(\hat n,\theta)\right)=e^{-i\theta\hat n\cdot\vec E_\ell}~,
\end{equation}
where the electric field $\vec E_\ell$ is the angular momentum conjugate to the $SU(2)$ rotor $h$ at link $\ell$. The bounded operator
\begin{equation}
H_\ell(f) = \int d\theta f(\theta)\int d\hat n 
e^{i\theta\hat n\cdot\vec E_\ell}~
\end{equation}
(where $f$ is any integrable function) is gauge-invariant. Indeed, it is a function of the gauge-invariant observable $\vec E_\ell^2$ that acts on the link $\ell$. 

As we have also explained in Sec.~\ref{sec:nondemolition}, the common-sense reason that the state created by the nonlocal operator in eq.~(\ref{correlated_flux_op}) cannot be created with local operators is that Alice's link and Bob's carry (correlated) nonzero electric charges. 
In quantum chromodynamics, as in a discrete gauge theory, Wilson loop measurement can cause color charge to be transferred to a flux tube. This color charge is surely detectable; like any color charge, it acts as a source for a measurable color electric field. 

\section{Matter fields and the destructive measurement of Wilson loops}
\label{sec:matter}
We have shown that the nondemolition measurement of a non-Abelian Wilson loop conflicts with relativistic causality. But there are further questions that we wish to address. Can the Wilson loop be measured {\em destructively}? What about the Abelian case? To formulate our answers, we will continue to use the formalism of lattice gauge theory. Furthermore, to ensure that the agents who are to perform measurements are as well equipped as possible, we will include in the theory matter fields that couple to the gauge fields.

Our matter fields reside on the sites of the lattice, and like the link variables, take values in the group algebra. The basis for the Hilbert space at a site $x$ will be denoted $\{|\phi\rangle_x, \phi\in G\}$. Under the local gauge transformation $U_x(g)$ acting at the site $x$, the matter variable transforms as the regular representation of $G$ (which contains all irreducible representations of $G$),
\begin{equation}
U_x(g):|\phi\rangle_x\to |g\phi\rangle_x~.
\label{regular_rep}
\end{equation} 
In addition to the gauge symmetry, the matter field at site $x$ also transforms under a {\em global} symmetry transformation $V_x(h)$, acting on $\phi$ from the right, that commutes with gauge transformations:
\begin{equation}
V_x(h): |\phi\rangle_x\to |\phi h^{-1}\rangle_x~.
\end{equation}
The interpretation of this global symmetry is that our matter fields have both  ``color'' and ``flavor'' degrees of freedom. The regular representation of $G$ decomposes into irreducible representations, with the dimension-$n_R$ representation $(R)$ occuring $n_R$ times. Gauge transformations mix the $n_R$ states that span $(R)$ (the colors), while global transformation mix the $n_R$ copies of $(R)$ (the flavors).

Let $xy$ denote a link connecting the neighboring sites $x$ and $y$ on the lattice, with orientation pointing from $y$ to $x$, and let $U_{xy}\in G$ be the gauge-field variable associated with this oriented link. We may also assign to this link the gauge-invariant variable 
\begin{equation}
u_{xy}=\phi_x^{-1} U_{xy}\phi_y~,
\end{equation}
the ``covariant derivative'' of the matter field. Without changing the physical content of the theory, we can replace the link variables $\{U_\ell\}$ by the new gauge-invariant variables $\{u_\ell\}$. But after this replacement, the physical Hilbert constraint can be trivially constructed: at each site $x$, the state of the matter field is required to be the gauge-invariant uniform superposition state
\begin{equation}
{1\over \sqrt{|G|}}\sum_{\phi\in G}|\phi\rangle_x~.
\end{equation}
Since the matter fields are completely constrained by gauge invariance, they have no role in dynamics and they too can be eliminated, leaving only the gauge-invariant local variables $\{u_\ell\}$. 

Although the new variables are gauge invariant, they  transform nontrivially under the global transformations, according to
\begin{eqnarray}
&&V_x(h): |u\rangle_{xy}\to |hu \rangle_{xy}~,\\
&&V_y(k): |u\rangle_{xy}\to |uk^{-1}\rangle_{xy}~.
\end{eqnarray}
Thus physical states can carry global $G$ charges.

A gauge-invariant unitary operator acting on the link $xy$ can be defined as
\begin{equation}
H_{xy}(a): |u\rangle_{xy}\to |au\rangle_{xy}~.
\end{equation}
Acting on the weakly-coupled vacuum, this operator produces a flux tube excitation at the link. The flux tube at $xy$ has Cheshire charge that is exactly compensated by charge localized at the site $x$. An operator $H_{xy}(\alpha)$ that creates an excitation with trivial Cheshire charge can be constructed and applied as described in Sec.~\ref{sec:flux}.

Since the variables $\{u_\ell\}$ are local and preserve the physical Hilbert space, it is reasonable to postulate that they are observable. Physically, the measurement of $u_{xy}$ has a simple interpretation in terms of the effect of parallel transport of a colored object from site $y$ to the neighboring site $x$. Of course, we are free to adopt arbitrary color conventions at each site, and the way we describe the effect of parallel transport depends on these conventions. However, if we have multiple flavors at our disposal, we can use the flavors to record our conventions, so that in effect (gauge-dependent) statements about color transport can be translated into (gauge-invariant) statements about flavors.

To be concrete, suppose that $(R)$ is a three-dimensional irreducible representation; our ``quarks'' come in three colors (red, yellow, blue) and three flavors (up, down, strange). At each site, we adopt conventions for color and flavor, and we prepare standard quarks in three mutually orthogonal colors and three mutually orthogonal flavors that lock these conventions together: the up quark is red, the down quark is yellow, the strange quark is blue. Then standard quarks prepared at site $y$ are covariantly transported to site $x$, and compared to the standard quarks that have been prepared at that site. Thus the effect of the transport can be equivalently described as either a rotation in the color space ($U_{xy}$) or in the flavor space ($u_{xy}$). Performing this experiment for each irreducible transformation of $G$ assigns a unique group element $u_{xy}$ to the link $xy$, for these particular conventions. A modification of the conventions can be interpreted as a rotation in the flavor space, under which the variable $u_{xy}$ transforms.

Now consider a large loop $C$ on the spatial lattice, and suppose that many parties distributed along the loop are to measure the Wilson loop 
\begin{equation}
W^{(R)}(C)\equiv\chi^{(R)}\left(\prod_{\ell\in C}U_\ell\right)=\chi^{(R)}\left(\prod_{\ell\in C}u_\ell\right)
\end{equation}
in representation $(R)$. Since all the matter fields cancel out, the Wilson loop can be expressed in terms of the gauge-invariant variables $\{u_\ell\}$. Each party has access to a single link along the loop, and using $n_R$ flavors of quarks in representation $(R)$, determines the value of the $n_R\times n_R$ matrix $D^{(R)}(u)$ at that link, for a particular choice of flavor conventions. Each party then reports her value of $D^{(R)}(u)$ to the central authority for post-processing, the matrices are multiplied together, and the trace is evaluated. The result, which does not depend on the local flavor conventions, is the value of the Wilson loop.

Thus, distributed parties, each acting locally, can measure the Wilson loop operator. But in doing so, they collect much additional information aside from the value of the Wilson loop. In particular, an eigenstate of $W^{(R)}(C)$ need not have a definite value of each $D^{(R)}(u_\ell)$ along the loop. Therefore, the localized measurement procedure typically disturbs the quantum state of the field, even if the initial state before the measurement is a Wilson loop eigenstate. Rather than a localized {\em nondemolition} measurement (which we have already seen is impossible) it is a localized {\em destructive} measurement. Note also that distributed parties can measure destructively each of several Wilson loop operators $W^{(R_i)}(C_i), i=1,2,3,\dots, n$, all on the same time slice, and hence the product $\prod_i W^{(R_i)}(C_i)$. In this respect, the destructive measurement is compatible with the Wilson loop algebra. Of course, we may not be able to measure more than one of the $W^{(R_i)}(C_i)$ if the $C_i$ are on different time slices, since a measurement on an earlier slice may interfere with a measurement on a later slice.

We have assumed that the matter fields transform as the regular representation eq.~(\ref{regular_rep}) of the gauge group $G$. What about more general choices for the representation content of the matter? Provided that the matter transforms as a faithful representation of the gauge group, one can show that the destructive measurement of $W^{(R)}(C)$ is still possible in any representation $(R)$.

We reach this conclusion by noting that matter in the regular representation can be {\em simulated} using matter that transforms faithfully, augmented by ancilla degrees of freedom. We will give only a brief sketch of the argument. First we recall that if $(R_m)$ is a faithful representation, and $(R)$ is any irreducible representation, then $(R)$ is contained in $(R_m)^{\otimes n}$ for some $n$. Therefore, if our fundamental matter fields transform as $(R_m)$, then we can build composite objects that transform as $(R)$ from $n$ fundamental constituents. 

Next we observe that if the theory contains only a single matter field that transforms as $(R)$, we can use ancilla variables to attach an effective flavor index to the field. To understand the point heuristically, consider the case of ``quarks'' that come in three colors but only one flavor. Rather than using ``natural'' flavors to keep track of our color conventions, we can use ``artificial'' flavors instead, labeling red, yellow, and blue quarks with the three mutually orthogonal states (up, down, strange) of the ancilla. When quarks are transported from one site to a neighboring site, the attached value of the ancilla is transported along with the color; hence artificial flavors, just like natural flavors, allow us to describe local gauge transport in terms of gauge-invariant quantities. Since we can construct composite matter fields in any representation $(R)$ of $G$, and we can use ancillas to ensure that matter transforming as $(R)$ comes in $n_R$ flavors, our simulated matter transforms as the regular representation;  thus we can measure destructively a Wilson loop in any representation.

What about the case of a pure gauge theory (one containing no charged matter at all)? The gauge variables themselves can simulate matter that transforms according to the adjoint representation
\begin{equation}
D(g):|h\rangle\to |ghg^{-1}\rangle~,
\end{equation}
which is a faithful representation of $G/Z(G)$, where $Z(G)$ denotes the center of $G$. Thus, by building composite fields and manipulating ancillas, we can simulate matter that transforms as the regular representation of $G/Z(G)$. Therefore, $W^{(R)}(C)$ can be measured destructively for any representation $(R)$ of $G/Z(G)$, or equivalently for any representation of $G$ that represents the center of $G$ trivially. 

We have seen that the nondemolition measurement of a non-Abelian Wilson loop is an example of an acausal measurement that can be made causal (and in fact localizable) if additional information is collected simultaneously. Other examples were noted in \cite{nielsen}.

\section{nondemolition measurement of Abelian Wilson loops is localizable}
\label{sec:abelian}

The causality problem arose for the nondemolition measurement of non-Abelian Wilson loops because multiplication of conjugacy classes is ill-defined. Since this problem does not arise if $G$ is Abelian, one might expect that a spacelike Wilson loop operator should be measurable in an Abelian gauge theory (or more generally, if the Wilson loop is evaluated in a one-dimensional irreducible representation of the gauge group). We will see that this is the case.

To be concrete, consider a lattice theory (containing charged matter) with gauge group $G=U(1)$. Gauge variables $U_\ell\in U(1)$ reside at each link $\ell$ of the lattice, and matter variables $\phi_x\in U(1)$ reside at each site $x$. As we have seen, the gauge and matter variables can be eliminated in favor of gauge-invariant variables $u_{xy}=\phi_x^{-1} U_{xy} \phi_y$, and the Wilson loop operator is 
\begin{equation}
W(C)\equiv \prod_{\ell\in C}U_\ell=\prod_{\ell\in C}u_\ell~.
\label{wilson_little_u}
\end{equation}
To perform a destructive measurement of $W(C)$, parties distributed along the loop $C$ could each measure the local value of $u$; then the results can be multiplied together later to determine the value of the Wilson loop.

To perform a nondemolition measurement of $W(C)$, the procedure must be modified so that only the value of the Wilson loop, and no further information, is collected. Imagine, then, that $n$ parties have been distributed along the loop $C$, each with access to one of the links of $C$. And suppose the party who resides at link $\ell$ can manipulate not only the gauge-invariant field variable $u_\ell$, but also a gauge-invariant {\em ancilla} variable $\tilde u_\ell\in U(1)$ that will be used to assist with the measurement. Some time ago, the parties prepared an entangled state of their ancilla variables, 
\begin{equation}
|{\rm initial}\rangle_{\rm anc} = \int \prod_{\ell=1}^n\left(d\tilde u_\ell\right)
|\tilde u_1,\tilde u_2, \dots, \tilde u_n\rangle \delta(\prod_{\ell=1}^n \tilde u_\ell - I)~.
\end{equation}
This state is a coherent superposition of all possible states for the ancilla variables, subject only to one global constraint on the product of all the $\tilde u_\ell$'s.
Now each party applies a local unitary transformation to her lattice field variable and her part of the ancilla:
\begin{equation}
\label{link_unitary}
|u_\ell,\tilde u_\ell\rangle \to |u_\ell, u_\ell \tilde u_\ell\rangle~, 
\end{equation}
a rotation of the ancilla rotor controlled by the value of the lattice rotor.
This is achieved by turning on a Hamiltonian that couples $u_\ell$ and $\tilde u_\ell$. 

The operation  eq.~(\ref{link_unitary}) modifies the constraint on the ancilla variables, which becomes
\begin{equation}
\label{new_constraint}
\prod_\ell \tilde u_\ell = W(C)~;
\end{equation}
Now each party can measure the value of her $\tilde u_\ell$, and broadcast the result to the central authority. The measurement outcomes are random, so that each individual measurement reveals no information about the state of the lattice variables. When the results are accumulated, the value of $W(C)$ can be inferred by evaluating $\prod_\ell \tilde u_\ell$, but no further information about the field configuration is acquired.  (This type of local measurement making use of a shared entangled ancilla was described in \cite{ahar86}, and was shown to be the basis of a separation between classical and quantum multiparty communication complexity in \cite{buhrman}.)

Of course, the transformation eq.~(\ref{link_unitary}) that couples the ancilla to the field variables can also be described in a conjugate basis, which may clarify its meaning. We may write $u=e^{-i\theta}$, $\tilde u=e^{-i\tilde \theta}$, and define the angular momentum $\tilde Q$ conjugate to $\tilde \theta$ by
\begin{equation}
e^{-i\tilde Q\tilde \xi}|\tilde \theta\rangle= |\tilde \theta+\tilde \xi\rangle~.
\end{equation}
Then eq.~(\ref{link_unitary}) becomes
\begin{equation}
|\theta,\tilde Q\rangle \to \left(e^{-i\theta}\right)^{\tilde Q}|\theta,\tilde Q\rangle~.
\label{charge_unitary}
\end{equation}
Thus we may regard $\tilde Q$ as a fictitious electric charge, whose transport properties are governed by the connection $u$ --- the parties implement eq.~(\ref{charge_unitary}) by ``parallel transporting'' their ancilla charges by one lattice spacing in the effective gauge field defined by $u$.
The (unnormalizable) initial state of the ancilla can be written
\begin{equation}
|{\rm initial}\rangle_{\rm anc}=\sum_{\tilde Q=-\infty}^\infty|\tilde Q,\tilde Q,\tilde Q,\dots,\tilde Q\rangle~,
\label{charge_sup_before}
\end{equation}
which is transformed to 
\begin{equation}
|{\rm initial}\rangle_{\rm anc}=\sum_{\tilde Q=-\infty}^\infty \left[W(C)\right]^{\tilde Q}|\tilde Q,\tilde Q,\tilde Q,\dots,\tilde Q\rangle~.
\label{charge_sup_after}
\end{equation}
Since the charges held by the parties are perfectly correlated, only the global information about transport around the entire loop $C$ becomes imprinted on the ancilla state. This information, encoded in relative phases in the $\tilde Q$-basis, can be read out via measurements in the conjugate $\tilde \theta$-basis. Note that it is important that the ancilla variables carry fictitious rather than genuine electric charges --- otherwise states with different values of the total charge would reside in distinct superselection sectors and the relative phases in eq.~(\ref{charge_sup_after}) would be unobservable. We also note that while to measure the Wilson loop perfectly we must prepare the ancilla in the unnormalizable (and hence unphysical) state eq.~(\ref{charge_sup_before}), a measurement with arbitrarily good precision can be achieved using a normalizable approximation to this state.  

The key to this procedure for measuring the Wilson loop is that the $W(C)$ can be expressed in terms of the local gauge-invariant variables $\{u_\ell\}$ as in eq.~(\ref{wilson_little_u}). This property has a clear physical interpretation. The matter field represents a medium laid out along the loop $C$ that becomes superconducting on the time slice where the measurement of $W(C)$ is to be carried out: $\phi=e^{-i\theta}$ is a superconducting order parameter with phase $\theta$. Though the phase and the gauge field $A_\mu$ are not locally observable, the covariant derivative
\begin{equation}
D_\mu\theta = \partial_\mu\theta + A_\mu
\end{equation}
is observable --- it is proportional to the local current density. By coupling the local current to our entangled ancilla, we have modified the state of the ancilla in a manner that is sensitive to the value of the quantity
\begin{equation}
\exp\left[i\oint_C D_\mu\theta dx^\mu\right]= W(C)~;
\end{equation}
the equality is obtained from the property that $\phi=e^{-i\theta}$ is a single-valued function.

Even without the entangled ancilla, parties distributed along the loop could determine the value of $W(C)$ by measuring the local value of $D_\mu \theta$, and broadcasting their results. In that case, not just $W(C)$ but also the covariant derivative of $\theta$ would be determined by their measurement outcomes. By invoking the entangled ancilla, we have emphasized that it is possible to measure $W(C)$ without learning anything else about the state of the lattice system, that is, to perform a nondemolition measurement of $W(C)$.

It is clear that the technique we have described could be applied in principle to perform a nondemolition measurement of the Wilson loop operator in any one-dimensional representation of the gauge group. But as we have shown must be so, it fails in the non-Abelian case. We can introduce matter fields such that $u_{xy}=\phi_x^{-1} U_{xy}\phi_y$ is a gauge-invariant quantity,  but since the $u$'s do not commute with the $\tilde u$'s, the transformation eq.~(\ref{link_unitary}) will not in that case simply modify the constraint on the ancilla variables as in eq.~(\ref{new_constraint}).

\section{Wilson loops in the pure Abelian gauge theory}
\label{sec:pure}

Our procedure for the nondemolition measurement of an Abelian Wilson loop uses charged matter coupled to the gauge fields. Let us now consider whether the nondemolition measurement is possible in the pure Abelian gauge theory. When there is no charged matter, we cannot replace the gauge variables on links by gauge-invariant variables that are locally measurable. 

\subsection{Homologically trivial loops}
Consider first the case of a homologically trivial loop $C$, the boundary of a two-dimensional surface $S$. In the Abelian gauge theory, the Wilson loop operator $W(C)$ can be interpreted as $e^{i\Phi}$ where $\Phi$ is the magnetic flux linking the loop. In the lattice formulation of the theory, the surface $S$ is the union of elementary cells that tessellate the surface. Suppose there are $N$ such cells, labeled by an index $\Sigma$ taking values $\Sigma=1,2,3,\dots,N$. Then the Wilson loop operator can be expressed as 
\begin{equation}
W(C) \equiv \prod_{\ell\in C} U_\ell=\prod_{\Sigma\in S} U_\Sigma~,
\end{equation}
where $U_\Sigma$ is the value of the Wilson operator $W(\partial \Sigma)$ for the boundary $\partial \Sigma$ of the cell $\Sigma$.
Therefore, a destructive measurement of $W(C)$ can be carried out by a collection of parties occupying the surface $S$. Each party measures the local ``magnetic field'' $U_\Sigma$ and reports her result to the central authority. The results can then be accumulated to determine the value of $W(C)$.

This destructive measurement differs from a nondemolition measurement of $W(C)$ in that too much information is collected --- not just the total flux through the surface, but also the local distribution of magnetic flux is determined by the measurement. In a nondemolition measurement of $W(C)$, a superposition of two different magnetic field configurations with the same value of $W(C)$ would not decohere, but if the local field is measured this superposition does decohere. Yet as in our previous discussion, a nondemolition measurement can be achieved if ancilla variables are prepared in an appropriate state that is distributed to the parties in advance. Suppose that each of the $N$ parties can access both the gauge-invariant local magnetic field variable $U_\Sigma\in U(1)$ and an ancilla variable $\tilde U_\Sigma\in U(1)$. The ancilla has been prepared in the shared initial state 
\begin{equation}
|{\rm init}\rangle_{\rm anc} = \int \prod_{\Sigma=1}^N\left(d\tilde U_\Sigma\right)
|\tilde U_1,\tilde U_2, \dots, \tilde U_N\rangle \delta(\prod_{\Sigma=1}^N \tilde U_\Sigma - I)~,
\end{equation}
a coherent superposition of all possible states for the ancilla variables, subject only to one global constraint on the product of all the $\tilde U_\Sigma$'s.
To perform the nondemolition measurement, each party applies a local unitary transformation to her magnetic flux variable and her part of the ancilla:
\begin{equation}
|U_\Sigma,\tilde U_\Sigma\rangle \to |U_\Sigma, U_\Sigma \tilde U_\Sigma\rangle~,
\end{equation}
a rotation of the ancilla rotor controlled by the value of the lattice rotor.
This operation modifies the constraint on the ancilla variables, which has
become
\begin{equation}
\prod_\Sigma \tilde U_\Sigma = \prod_\Sigma U_\Sigma = W(C)~.
\end{equation}
Now each party can measure the value of her $\tilde U_\Sigma$, and broadcast the result to the central authority. The measurement outcomes are random, so that each individual measurement reveals no information about the state of the lattice variables. When the results are accumulated, the value of $W(C)$ can be inferred by evaluating $\prod_\Sigma \tilde U_\Sigma$, but no further information about the gauge field configuration is acquired.  

\subsection{Homologically nontrivial loops}

Now consider the case of a homologically nontrivial loop $C$, which is not the boundary of any surface. For example, suppose that the theory lives on a $d$-dimensional spatial torus (a rectangular box with opposite sides identified), and that the loop $C$ is a nontrivial cycle that winds around the torus. 

The gauge-invariant local operators of the theory are the magnetic flux operators $U_\Sigma$ acting on the elementary lattice cells, and the ``electric field'' operators that act on elementary links. The electric field $E_\ell$ at the link $\ell$ is the ``angular momentum'' conjugate to the link rotor variable $U_\ell$; it generates rotations of $U_\ell$
\begin{equation}
\exp(-i\theta E_\ell): |U\rangle_\ell \to |e^{-i\theta}U\rangle_\ell~.
\end{equation}
Each party residing on the lattice is empowered to apply or measure the local operators in her vicinity.

But the homologically nontrivial Wilson loop operator is not included in the algebra generated by these local operations. Hence $W(C)$, where $C$ is a nontrivial cycle, is completely inaccessible to the local residents of the lattice. They cannot measure this operator, either destructively nor nondestructively, nor can they apply it to a state. The homologically nontrivial Wilson loop is not an observable of the pure gauge theory.

Although the inhabitants of this world are unable to measure $W(C)$, they are able to change its value. The link rotation $e^{-i\theta E_\ell}$ has a nontrivial commutation relation with $W(C)$ if $\ell\in C$:
\begin{equation}
e^{-i\theta E_\ell} W(C) = e^{-i\theta} W(C) e^{-i\theta E_\ell}~.
\end{equation}
(Here the orientation of the link $\ell$ used to define $E_\ell$ is assumed to be aligned with the orientation of $C$ at link $\ell$.) Thus any party with access to a link $\ell$ of $C$ can rotate the value of $W(C)$, whether or not $W(C)$ is the boundary of a surface. 

\begin{figure}
\begin{center}
\leavevmode
\epsfxsize=3in
\epsfbox{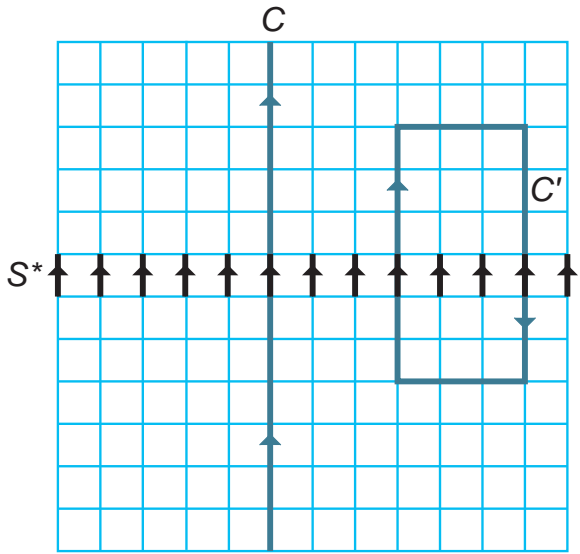}
\end{center}
\caption{The nonlocal electric field operator dual to a homologically nontrivial Wilson loop operator $W(C)$, in an Abelian lattice gauge theory in two spatial dimensions. Here a two-torus is represented as a square with opposite sides identified, $C$ is a nontrivial oriented cycle that winds around the torus, and $S^*$ is the set of oriented links dual to a closed ``surface'' that crosses $C$ once.  Any homologically trivial closed loop (like $C'$) crosses $S^*$ as many times with a $+$ orientation as with a $-$ orientation. Thus the electric field operator on $S^*$ commutes with $W(C')$, but has a nontrivial commutation relation with $W(C)$.}
\label{fig:dualsurface}
\end{figure}

Like Wilson loop operators, electric field operators are of two types with differing locality properties. If $C$ is a fundamental nontrivial cycle, we can construct an electric field operator $E_C$ that rotates $W(C)$ but has no effect on homologically trivial Wilson loops. Associated with the cycle $C$ of the torus is a closed orientable hypersurface $S$ that crosses $C$ exactly once; dual to this surface is a set of oriented lattice links $S^*$, as illustrated in Fig.~\ref{fig:dualsurface}. The electric field conjugate to $W(C)$ is 
\begin{equation}
E_C=\sum_{\ell\in S^*} E_\ell~.
\end{equation}
This nonlocal operator generates a rotation of the homologically nontrivial Wilson loop $W(C)$, but since any homologically trivial closed loop crosses $S$ as many times with a $+$ orientation as with a $-$ orientation, homologically trivial Wilson loop operators commute with $E_C$.

The ``nonlocal electric field'' $E_C$ can be measured --- all parties residing at links contained in $S^*$ can measure the local electric field and the results can be summed. But while the inhabitants of the lattice are able to measure $E_C$, they are unable to change its value. The Hilbert space of the theory divides into superselection sectors, each labeled by the values of $E_{C_i}\in Z$, where the $C_i$'s are the cycles that generate the homology group of the spatial manifold.

It is obvious that similar conclusions apply to any Abelian pure gauge theory. If the theory is defined on a manifold with nontrivial homology, then the algebra of observables has a different structure in the theory with charged matter than in the pure gauge theory without matter. In the pure gauge theory, the homologically nontrivial Wilson loops are not observables at all, and consequently, the theory divides into sectors with different values of the nonlocal electric field.

\section{Superluminal charge transport}
\label{sec:sctl}

The main conclusion of this paper is that the observables of Abelian and non-Abelian gauge theories have fundamentally different properties --- in particular, the nondemolition measurement of a Wilson loop is acasual in the non-Abelian case and localizable in the Abelian case. We can further appreciate the distinction between Abelian and non-Abelian gauge theories by thinking about, not what operators can be {\em measured}, but rather what operators can be {\em applied} to a state by a group of parties each of whom acts locally.

To dramatize the question, imagine two parties Alice and Bob, many light years apart, who share a ``superluminal charge transport line'' (SCTL). Alice places a single electrically charged particle, an electron, at her end of the SCTL (the point $y$); then her charge mysteriously disappears, and in an instant reappears at Bob's end of the SCTL (the point $x$). The electron has been transmitted through the SCTL far more rapidly than Alice could send a light signal to Bob. Is such a device physically possible?

Yes. We can understand how the SCTL works by characterizing it with a gauge-invariant unitary operator that it applies to a state. In our lattice formulation of an Abelian lattice gauge theory with matter, consider a connected path of links $P$ that begins at $y$ and ends at $x$. Associated with this path is the gauge-invariant operator
\begin{equation}
\label{openpathproduct}
\phi^{-1}_x\left(\prod_{\ell\in P}U_\ell\right)\phi_y=\prod_{\ell\in P}u_\ell~.
\end{equation}
Acting on the weakly-coupled ground state of the theory, this operator creates a pair of equal and opposite charges at the sites $x$ and $y$. Acting on a state with a charged particle at site $y$, it annihilates the particle at $y$ while creating a particle of like charge at $x$, in effect transporting the particle from $y$ to $x$. The applied operator factorizes as a product of gauge-invariant unitary operators $u_\ell$, each acting on a single lattice link. Therefore, many parties acting simultaneously, each manipulating only the link in her own vicinity, are able to operate the SCTL.

\begin{figure}[t]
\begin{center}
\leavevmode
\epsfxsize=3in
\epsfxsize=3in
\epsfbox{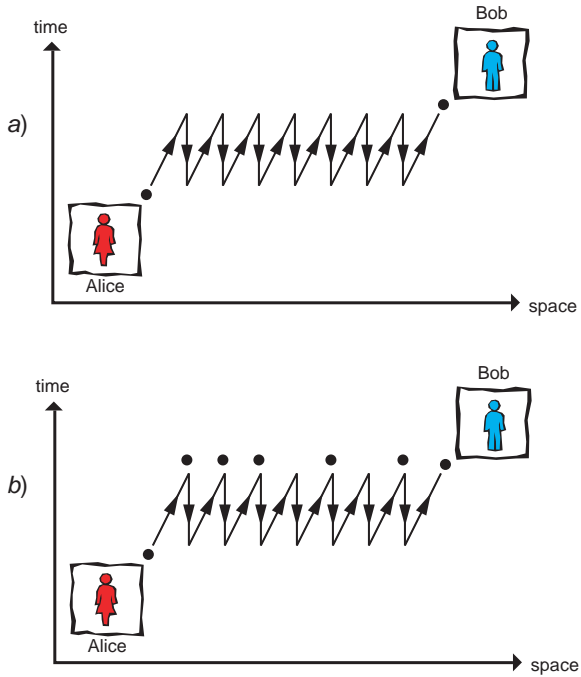}
\end{center}
\caption{$(a)$ ``Sawtooth protocol'' for superluminal transmission of an electron from Alice to Bob, assisted by many intervening parties. Each party (except Alice and Bob) produces an electron-positron pair and keeps the positron, and each (except Bob) passes an electron to the party on her right. Then all pairs annihilate.  Thus a charged particle sent by Alice is received by Bob almost instantaneously, even though Bob is many light years away. $(b)$ The protocol fails to achieve superluminal transport of non-Abelian charge. All intervening parties produce color-singlet pairs of charges, but when each party unites her antiparticle with the particle created by her neighbor, the pairs fail to annihilate completely. Though the procedure conserves color, the color of the charge received by Bob is uncorrelated with the color of the particle that had been in Alice's possession. In both the Abelian and non-Abelian cases, no information is transmitted from Alice to Bob, so that causality is not violated.}
\label{fig:sawtooth}
\end{figure}

More physically, we can envision the operation of the SCTL as in Fig.~\ref{fig:sawtooth}. Many parties are distributed along the SCTL. At a pre-arranged time, each party creates an electron-positron pair. Retaining the positron, she passes the electron to her right, while receiving an electron from the party on her left. Then she brings electron and positron together to annihilate. Claire, the party closest to Alice, receives an electron from Alice and annihilates it with Claire's positron, while Diane, the party closest to Bob, hands her electron to Bob. After all pairs annihilate, the sole remaining electron, initially in Alice's hands, has been delivered to Bob. The closer the parties are to one another, the faster the procedure can be completed.

Even though the charge transfer is virtually instantaneous, the Gauss law is satisfied at all times. If we draw surfaces around Alice and Bob, then while the SCTL is operating one unit of charge leaves Alice's surface and one unit enters Bob's. Furthermore, even though the charge moves superluminally, the process does not violate causality, since no information is transmitted from Alice to Bob. Indeed, if Felicity, a party in the middle of the SCTL, were to disobey orders and fail to create an electron-positron pair, then Felicity would ``intercept'' the charge sent by Alice, and Felicity's neighbor on the right would share a distantly separated electron-positron pair with Bob. When Bob receives the electron, all he learns is that his left neighbor has performed as expected, but he learns nothing about the activities of Alice. 

While an Abelian charge really carries no information, non-Abelian charge is much more interesting --- its orientation in a representation space can encode a message. Thus it is easy to see that a non-Abelian SCTL, were one to exist, would violate causality. To be explicit, consider the following protocol that enables Bob to send classical information to Alice (based on ideas similar to those used to show that Wilson loop measurement is acausal). First, Alice produces a particle-antiparticle pair, where the particle transforms as representation $(R)$ of $G$ and the antiparticle as representation $(R^*)$. The total charge of the pair is trivial. If $\{|e_i\rangle, i=1,2,3,\dots n_R\}$ denotes a basis for the representation $(R)$, and $\{|e_i^*\rangle\}$ denotes the conjugate basis for $(R^*)$, then the singlet state prepared by Alice is
\begin{equation}
{1\over\sqrt{n_R}}\sum_i|e_i\rangle\otimes |e_i^*\rangle~.
\end{equation}
Alice keeps the antiparticle, and sends the particle through the SCTL to Bob. Bob has a loop of magnetic flux that he has prepared in the charge-zero state $|\alpha,0\rangle$ associated with conjugacy class $\alpha$ of $G$, as in eq.~(\ref{chargezero_alpha}). To convey a bit of information to Alice, Bob either does nothing to the charged particle he received from Alice (sending $0$) or lassoes it with his flux tube (sending $1$), and returns the charge through the SCTL to Alice. Now if Bob did nothing, Alice recovers a singlet pair, but if Bob lassoed the charge, then the state of the pair has become entangled with the state of Bob's tube:
\begin{equation}
\label{lassoed_charge}
{1\over\sqrt{n_R}}{1\over\sqrt{|\alpha|}}\sum_{a\in\alpha}\sum_{i,j}|e_i\rangle\otimes |e_j^*\rangle D_{ij}(a)\otimes |a\rangle~.
\end{equation}
Alice then unites the particles and observes whether the pair annihilates. In the state eq.~(\ref{lassoed_charge}), the probability of annihilation is determined by the overlap of the pair's state with the singlet state, and is readily seen to be
\begin{equation}
{\rm Prob}= \left|{1\over n_R}\chi^{(R)}(\alpha)\right|^2~,
\end{equation}
where $\chi^{(R)}(\alpha)$ is the character of class $\alpha$ in representation $(R)$. As long as the representation $(R)$ is not one-dimensional, the class $\alpha$ can be chosen so that this probability is less than one. Therefore, Alice observes annihilation with certainty if Bob sends 0 and observes annihilation with probability less than unity if Bob sends 1 --- thus Bob can signal Alice.

The capacity of the SCTL is easily estimated. Suppose that Alice will signal Bob by transmitting $N$ particles (where $N$ is even) each transforming as the representation $(R)$ or its conjugate representation $(R^*)$. She can prepare and send a state of $N/2$ particles and $N/2$ antiparticles, in any one of $A_N$ distinct singlet states. These states are mutually orthogonal and in principle they can be readily distinguished by Bob. Therefore, Alice is able to send $\log_2 A_N$ bits to Bob by using the SCTL $N$ times. But the number of singlets is
\begin{equation}
A_N= {\left(n_R\right)^N\over P(N)}~,
\end{equation}
where $P(N)$ grows no faster than a polynomial with $N$. Thus, asymptotically Alice can send $\log_2 n_R$ bits of information per transmission. This rate is just what we would have guessed naively, ignoring that observables must be gauge invariant.

Since the non-Abelian SCTL is acausal, it ought not to be physically realizable. What goes wrong if we try the same procedure that succeeded in the Abelian case? The trouble is that if Claire produces a singlet pair, and Diane does the same, then when Claire's particle unites with Diane's antiparticle, the charges might be unable to annihilate. In fact, if Claire's particle transforms as the representation $(R)$ and Diane's as $(R^*)$, then the probability that the pair annihilates, determined by its overlap with the single state, is $1/n_R^2$. Thus, while in the Abelian case the outcome of the procedure is that only a single electron survives, which is in Bob's possession, in the non-Abelian case many relic charges remain strewn along the path of the would-be SCTL. Though the procedure conserves charge, the orientation in the representation space of the charge that Bob receives is actually uncorrelated with the orientation of the charge that Alice sent, and no information is transmitted.

Finally, in the non-Abelian theory as in the Abelian theory, the operator that propagates a charged particle from $y$ to $x$ can be factorized as in eq.~(\ref{openpathproduct}) into local factors. So why can't this operator be applied by many parties, each acting locally? We must recall that the operators of the theory are not the group elements $u_\ell\in G$ themselves, but rather the matrix elements $D^{(R)}_{ij}(u_\ell)$ of representations of the group. In the Abelian case, the character of a product of group elements can be written as a product of characters, where each character is a unitary operator. But in the non-Abelian case, the ``factorized'' operator is really a product of matrices. The contraction of indices in this matrix product cannot be achieved by many  parties acting locally; rather it requires a nonlocal conspiracy.

\section{Conclusions}
\label{sec:conclusions}

In the standard formulation of algebraic relativistic quantum field theory \cite{haag}, an algebra of ``local'' operators on Hilbert space is associated with each bounded open region of spacetime, such that two local operators commute if they are associated with regions that are spacelike separated. A local operator is designated as a ``local observable'' if it preserves the superselection sectors of the theory. One might be tempted to postulate that a quantum operation is physically possible in principle if and only if it can be expanded in terms of these local observables. 

We find this viewpoint untenable, because causality places  more stringent constraints on the allowed operations \cite{nielsen}. The problem of characterizing which quantum operations are compatible with causality is especially subtle, interesting, and physically relevant in relativistic quantum field theories with local gauge symmetry. 

One form of the question is: what operators can justifiably be called ``observables?'' We have focused our attention on the measurability of the Wilson loop because of its prominent place in the operator algebra of a gauge theory. The answer we have found is rather elaborate. In a gauge theory that includes charged matter that transforms faithfully, a {\em destructive} measurement of a spacelike Wilson loop $W^{(R)}(C)$ is physically possible for any representation $(R)$ of the gauge group. The term ``destructive'' means that many cooperating parties acting together can ascertain the value of the Wilson loop, but only by collecting additional information in the process, and at the price of damaging Wilson loop eigenstates. In a pure gauge theory (one with no charged matter), the destructive measurement of $W^{(R)}(C)$ is possible for any $(R)$ that represents the center of the gauge group trivially. A nondemolition measurement of the Wilson loop (one that leaves Wilson loop eigenstates intact) is possible in an Abelian gauge theory but not in a non-Abelian gauge theory. 

Nondemolition measurement of a non-Abelian Wilson loop is impossible because it would conflict with relativistic causality. Two distantly separated parties (Alice and Bob), can each produce excitations locally (magnetic flux tubes), preparing a state that is not an eigenstate of the Wilson loop operator $W(C)$, where $C$ is a loop that passes through both excitations. Projecting onto a Wilson loop eigenstate, whatever the outcome, entangles Alice's excitation with Bob's, modifying the excitations in a manner that either party can discern locally. Such instantaneous preparation of quantum entanglement would enable spacelike-separated Alice and Bob to communicate.

In quantum field theory in general, and in gauge theories in particular, characterizing the physically allowed quantum operations seems to be an open problem. Further progress on this question is bound to elucidate the physical content of relativistic quantum theory.

\acknowledgments

We thank Steve Giddings, Anton Kapustin, Michael Nielsen, Edward Witten, and especially Mark Srednicki for helpful discussions and comments. This work was supported in part by the Department of Energy under Grant No. DE-FG03-92-ER40701, by the National Science Foundation under Grant No. EIA-0086038, by the Caltech MURI Center for Quantum Networks under ARO Grant No. DAAD19-00-1-0374, and by an IBM Faculty Partnership Award. Some of this work was done at the Aspen Center for Physics. This research was conducted in part while D.G. served
as a Clay Long-Term CMI Prize Fellow.


\end{document}